\documentclass[runningheads]{svmult}

\usepackage{makeidx}   
\usepackage{graphicx}  
\usepackage{subeqnar}  
\usepackage{multicol}  
\usepackage{physmubb}  
\makeindex             

\begin{document}
%
\topmargin3cm
%
\title*{Hadron Colliders, the Standard Model,\\and Beyond}
\toctitle{Hadron Colliders, the Standard Model, and Beyond}
%
%
\titlerunning{Hadron Colliders, the Standard Model, and Beyond}
%
\author{Scott Willenbrock \\ \bigskip\smallskip
{\it Department of Physics, University of Illinois at Urbana-Champaign, \\
1110 West Green Street, Urbana, IL   61801 \\ \medskip Theoretical Physics
Department, Fermi National Accelerator Laboratory, P.~O.~Box 500, Batavia, IL
60510}}
\authorrunning{S.~Willenbrock}
%
%

\maketitle              

\section{What is the Standard Model?}
\label{sec:sm}

Quantum field theory combines the two great achievements of 20$^{\rm
th}$-century physics, quantum mechanics and relativity.  The standard model is
a particular quantum field theory, based on the set of fields displayed in
Table~\ref{tab:sm}, and the gauge symmetries $SU(3)\times SU(2)\times U(1)_Y$.
There are three generations of quarks and leptons, labeled by the index
$i=1,2,3$, and one Higgs field, $\phi$.

\begin{table}[b]
\caption{The fields of the standard model and their gauge quantum numbers.}
\begin{center}\begin{tabular}[7]{cccccccc}
&&&&$\underline{SU(3)}$&$\underline{SU(2)}$&$\underline{U(1)_Y}$\\
\\
$Q_L^i=$&$\left(\begin{array}{l}u_L\\d_L\end{array}\right)$
&$\left(\begin{array}{l}c_L\\s_L\end{array}\right)$
&$\left(\begin{array}{l}t_L\\b_L\end{array}\right)$&3&2&$\frac{1}{6}$\\
\\
$u_R^i=$&$u_R$&$c_R$&$t_R$&3&1&$\frac{2}{3}$\\
\\
$d_R^i=$&$d_R$&$s_R$&$b_R$&3&1&$-\frac{1}{3}$\\
\\
$L_L^i=$&$\left(\begin{array}{l}\nu_{eL}\\e_L\end{array}\right)$
&$\left(\begin{array}{l}\nu_{\mu L}\\\mu_L\end{array}\right)$
&$\left(\begin{array}{l}\nu_{\tau
L}\\\tau_L\end{array}\right)$&1&2&$-\frac{1}{2}$\\
\\
$e_R^i=$&$e_R$&$\mu_R$&$\tau_R$&1&1&$-1$\\ \\
$\phi=$&$\left(\begin{array}{l}\phi^+\\\phi^0\end{array}\right)$
&&&1&2&$\frac{1}{2}$
\end{tabular}\end{center} \label{tab:sm}
\end{table}

Once the fields and gauge symmetries are specified, the standard model is the
most general theory that can be constructed.  The only constraint imposed is
that the interactions in the Lagrangian be the simplest possible.  We will
return to this point in Section~\ref{sec:numasses}

Let's break the Lagrangian of the standard model into pieces:
\begin{equation}
{\cal L}_{SM}={\cal L}_{Gauge}+{\cal L}_{Matter}+{\cal L}_{Yukawa}+{\cal
L}_{Higgs}\;.
\end{equation}
The first piece is the pure gauge Lagrangian, given by
\begin{equation}
{\cal L}_{Gauge} = \frac{1}{2g_S^2}{\rm Tr}\;
G^{\mu\nu}G_{\mu\nu}+\frac{1}{2g^2}{\rm Tr}\;
W^{\mu\nu}W_{\mu\nu}-\frac{1}{4}B^{\mu\nu}B_{\mu\nu}\;, \label{Lgauge}
\end{equation}
where $G^{\mu\nu}$, $W^{\mu\nu}$, and $B^{\mu\nu}$ are the gluon, weak, and
hypercharge field-strength tensors.  These terms contain the kinetic energy of
the gauge fields and their self interactions.  The next piece is the matter
Lagrangian, given by
\begin{equation}
{\cal L}_{Matter}=i\bar Q_L^i\not\!\!DQ_L^i+i\bar u_R^i\not\!\!Du_R^i+i\bar
d_R^i\not\!\!Dd_R^i+i\bar L_L^i\not\!\!DL_L^i+i\bar e_R^i\not\!\!De_R^i\;.
\label{Lmatter}
\end{equation}
This piece contains the kinetic energy of the fermions and their interactions
with the gauge fields, which are contained in the covariant derivatives. These
two pieces of the Lagrangian depend only on the gauge couplings $g_S, g, g'$.
Mass terms for the gauge bosons and the fermions are forbidden by the gauge
symmetries.

The next piece of the Lagrangian is the Yukawa interaction of the Higgs field
with the fermions, given by
\begin{equation}
{\cal L}_{Yukawa} = -\Gamma_u^{ij}\bar Q_L^i\epsilon
\phi^*u_R^j-\Gamma_d^{ij}\bar Q_L^i\phi d_R^j-\Gamma_e^{ij}\bar L_L^i\phi
e_R^j + h.c.\;, \label{LYukawa}
\end{equation}
where the coefficients $\Gamma_u$, $\Gamma_d$, $\Gamma_e$ are $3\times 3$
complex matrices in generation space.  They need not be diagonal, so in
general there is mixing between different generations.  These matrices contain
most of the parameters of the standard model.

\begin{figure}
\begin{center}
\includegraphics[width=.5\textwidth]{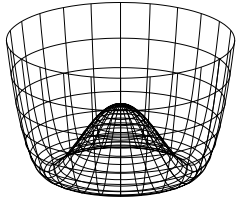}
\end{center}
\caption[]{The Higgs potential.  The neutral component of the Higgs field
acquires a vacuum-expectation value $\langle \phi^0\rangle = v/\sqrt 2$ on the
circle of minima in Higgs-field space.} \label{higgspotential}
\end{figure}

The final piece is the Higgs Lagrangian, given by
\begin{equation}
{\cal L}_{Higgs} = (D^\mu\phi)^\dagger
D_\mu\phi+\mu^2\phi^\dagger\phi-\lambda(\phi^\dagger\phi)^2\;.\label{LHiggs}
\end{equation}
This piece contains the kinetic energy of the Higgs field, its gauge
interactions, and the Higgs potential, shown in Fig.~\ref{higgspotential}. The
coefficient of the quadratic term, $\mu^2$, is the {\em only} dimensionful
parameter in the standard model.  The sign of this term is chosen such that
the Higgs field has a nonzero vacuum-expectation value on the circle of minima
in Higgs-field space given by $\langle\phi^0\rangle=\mu/\sqrt{2\lambda} \equiv
v/\sqrt 2$. The dimensionful parameter $\mu$ is replaced by the dimensionful
parameter $v\approx 246$ GeV.

The acquisition of a nonzero vacuum-expectation value by the Higgs field
breaks the electroweak symmetry and generates masses for the gauge bosons,
\begin{eqnarray}
M_W=\frac{1}{2}gv\nonumber\\
M_Z=\frac{1}{2}\sqrt{g^2+g'^2}\,v\;, \label{mw}
\end{eqnarray} and the fermions,
\begin{equation}
M=\Gamma \frac{v}{\sqrt 2}\;.
\end{equation}
Diagonalizing the fermion mass matrices generates the
Cabibbo-Kobayashi-Maskawa (CKM) matrix, including the CP-violating phase.

This concludes my lightning review of the standard model.\footnote{For a less
frenetic review, see Ref.~\cite{Willenbrock:2002ta}.} It is impressive how
tight the structure is; once the fields and the gauge symmetries are
specified, the rest follows automatically.

This answers the question posed in the title of this section, but the more
important question is: Is the standard model correct?  I show in Fig.~2 a
qualitative assessment of the pieces of the standard model described above.
The gauge-boson self interactions and the gauge interactions of the fermions
(with the exception of the top quark) have been tested to very good accuracy,
so we are sure that they are described by the standard model.  The top-quark's
gauge interactions have been tested less accurately, but thus far they agree
with the standard model.  Quark mixing agrees with the CKM picture to very good
accuracy in the first two generations, but with less accuracy in the third
generation.  The relation between the $W$- and $Z$-boson masses that follows
from Eq.~(\ref{mw}),
\begin{equation}
M_W^2=M_Z^2\cos^2\theta_W(1+\Delta\rho)\;,
\end{equation}
where $\Delta\rho$ contains the radiative corrections, has been tested to good
accuracy, but there are plans to do even better (see Section~\ref{sec:ew}). The
pieces of the standard model that we have no direct knowledge of involve the
coupling of the Higgs boson to fermions, gauge bosons, and to itself.

\begin{figure}
\begin{center}
\begin{math}
\begin{array}{rclcl}
\begin{array}{r} \bf{\cal{L}_{\it{Gauge}}:} \\
                 \vspace{.42in} \end{array} &
\hspace{.1in} & \includegraphics[width=.4\textwidth]{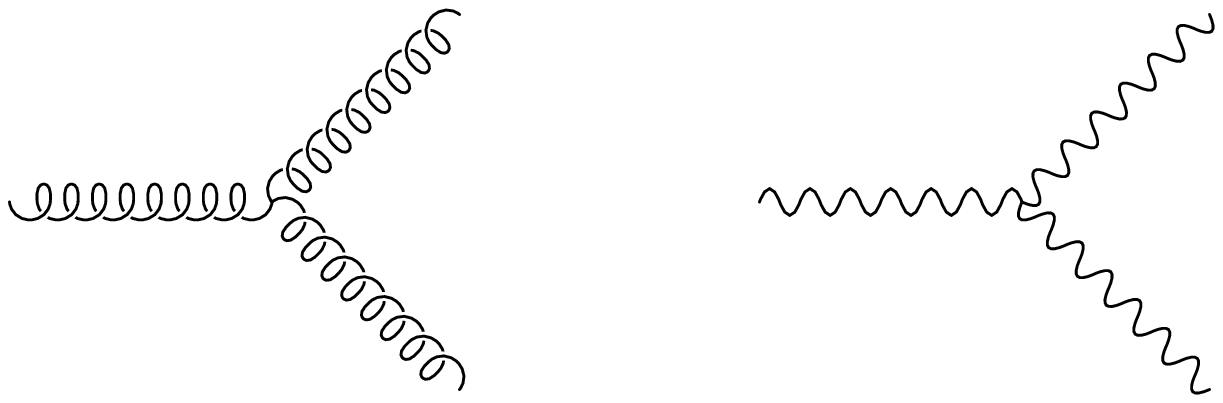} & \hspace{.1in}
&
\begin{array}{l} \rm{Yes} \\
                 \vspace{.42in} \end{array} \\
\begin{array}{r} \bf{\cal{L}_{\it{Matter}}:} \\
                 \vspace{.37in} \end{array} &
\hspace{.1in} & \includegraphics[width=.4\textwidth]{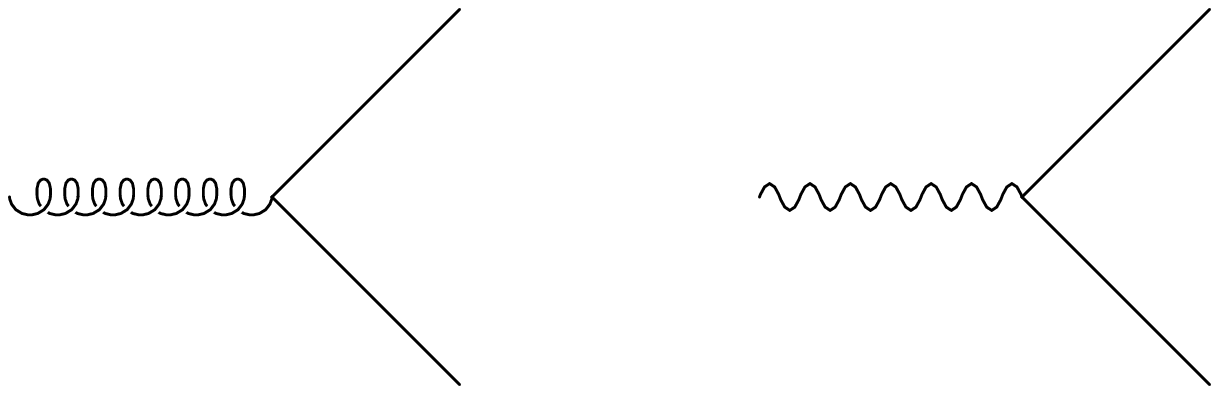} &
\hspace{.1in} &
\begin{array}{l} \rm{Yes \;\; (u,d,s,c,b)} \\
                 \rm{Probably \;\; (t)} \\
                 \vspace{.32in} \end{array} \\
\bf{\cal{L}_{\it{Yukawa}}:} & \hspace{.1in} & \;\; \rm{CKM \; mixing} &
\hspace{.1in} &
\begin{array}{l} \rm{Yes \;\; (u,d,s,c)} \\
                 \rm{Probably \;\; (b,t)} \end{array} \\[0.4in]
& \hspace{.1in} & \includegraphics[width=.17\textwidth]{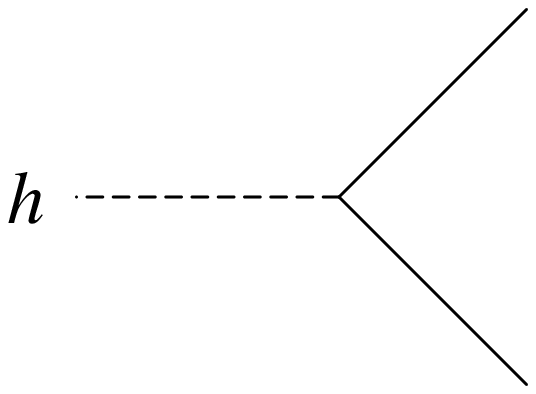} &
\hspace{.1in} &
\begin{array}{l} \rm{Don't \; know} \\
                 \vspace{.35in} \end{array} \\
\bf{\cal{L}_{\it{Higgs}}:} & \hspace{.1in} & {M_{W}^{2} = M_{Z}^{2} \, \cos^2
\theta_W \; (1+\Delta\rho)} &
\hspace{.1in} & \rm{Probably} \\[0.4in]
& \hspace{.1in} & \includegraphics[width=.4\textwidth]{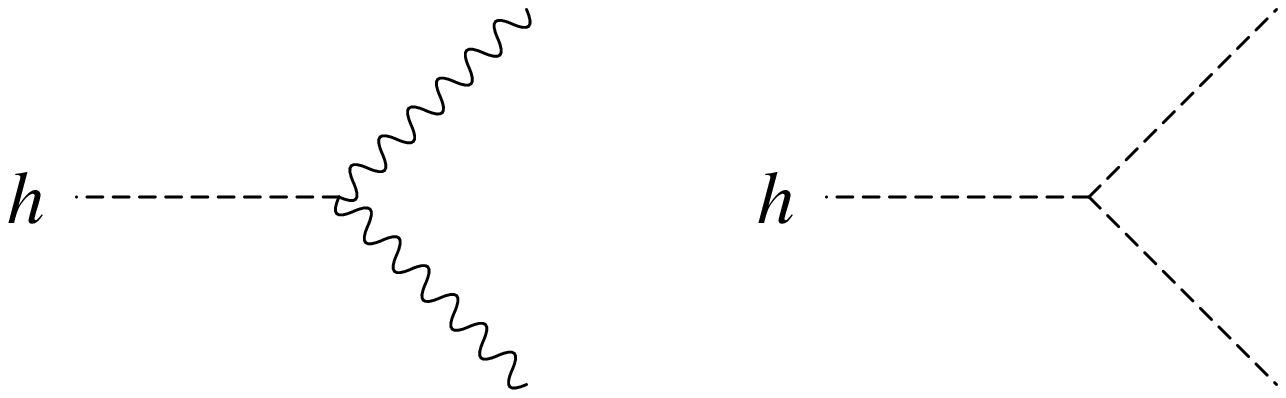} &
\hspace{.1in} &
\begin{array}{l} \rm{Don't \; know} \\
                 \vspace{.4in} \end{array}
\end{array}
\end{math}
\end{center}\vspace{-.25in}\caption{A qualitative answer to the question:
Is the standard model correct?}\label{smassessment}
\end{figure}

I argue in the next section that the Fermilab Tevatron will contribute to our
knowledge of the pieces of the standard model that are listed as ``probably''
correct in Fig.~\ref{smassessment}. If we are fortunate, it will also
contribute to our knowledge of the Higgs sector.  The CERN Large Hadron
Collider (LHC) will certainly discover the Higgs boson, as well as measure its
coupling to other particles.

\section{Hadron Colliders and the Standard Model}

The LHC was designed to discover the Higgs boson.  The question I would like
to address in this section is: How can the Tevatron confront the standard
model?  I discuss five of the most important ways in which it can do so.
Please regard these as appetizers for the more detailed talks that will follow
at this conference.

\newpage

\subsection{Precision electroweak}
\label{sec:ew}

\begin{figure}[t]
\begin{center}
\includegraphics[width=.6\textwidth]{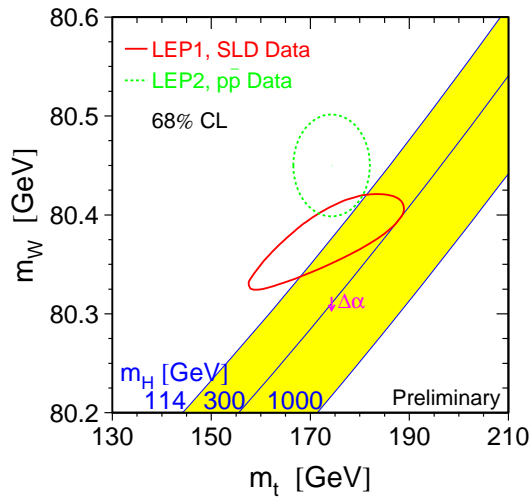}
\end{center}
\caption{Lines of constant Higgs mass on a plot of $M_W$ {\it vs.}~$m_t$.  The
dashed ellipse is the $68\%$ CL direct measurement of $M_W$ and $m_t$.  The
solid ellipse is the $68\%$ CL indirect measurement from precision electroweak
data.  From http://lepewwg.web.cern.ch/LEPEWWG.} \label{mwmt}
\end{figure}

I show in Fig.~\ref{mwmt} a plot of $M_W$ {\it vs.} $m_t$, with lines of
constant Higgs mass.  The dashed ellipse is the $68\%$ CL region from direct
measurements; the solid ellipse is the $68\%$ CL region from indirect
measurements.  The fact that these ellipses lie near each other, and near the
lines of constant Higgs mass (for $m_h>114$ GeV) tells us that the standard
model is not obviously wrong.  It also indicates that the Higgs boson is
light, $m_h<204$ GeV at $95\%$ CL \cite{Hagiwara:pw}.

Before we become too smug about the success of the standard model in precision
electroweak analyses, note that the fit to all data has a CL of 0.01, hardly
confidence inspiring \cite{Chanowitz:2002cd}.  This is due to two anomalous
measurements: the $b$ forward-backward asymmetry ($A_{FB}^b$) measured at LEP,
which deviates by $2.6\sigma$, and $\sin^2\theta_W(\nu N)$ measured by the
NuTeV collaboration \cite{Zeller:2001hh}, which deviates by $3\sigma$.  What
if we (very unscientifically) discard these measurements from the global fit?
Then we find that $m_h> 114$ GeV has a CL of only 0.03, because these two
measurements favor a heavy Higgs boson, while the other measurements favor a
very light Higgs boson.  If we only discard the NuTeV measurement, then the
fit CL is 0.10, which is marginally acceptable. It seems that there is some
tension in the fit of the precision electroweak data to the standard model.

Measurements of $M_W$ and $m_t$ at the Tevatron could resolve or exacerbate
this tension.  Let us take as a goal for Run IIa (2 fb$^{-1}$) uncertainties of
$\Delta M_W=30$ MeV, $\Delta m_t=3$ GeV.  Combined with the LEP measurement of
$\Delta M_W=42$ MeV, the uncertainty in the $W$ mass would be $\Delta M_W=24$
MeV.  The goals for Run IIb are $\Delta M_W= 20$ MeV, $\Delta m_t=$ 2 GeV.
These measurements will be important tests of the standard model, either
increasing our confidence in it or indicating that there is physics beyond it.

\begin{figure}[t]
\begin{center}
\includegraphics[width=\textwidth]{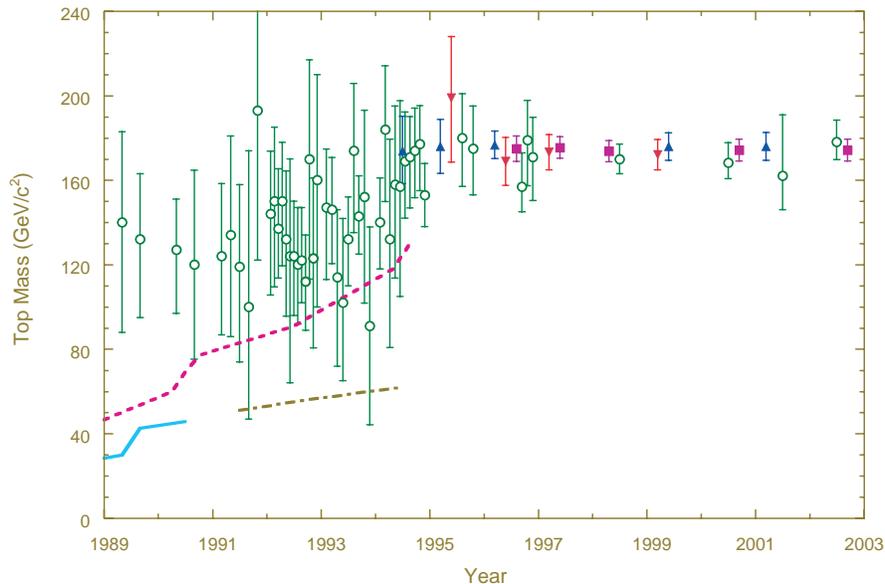}
\end{center}
\caption{($\diamondsuit$) Indirect bounds on the top-quark mass from precision
electroweak data. ($\sqcap\!\!\!\!\sqcup$) World-average direct measurement of
the top-quark mass; ($\bigtriangleup$) CDF and ($\bigtriangledown$) D0
measurements.  Lower bounds from $p\bar p$ (dashed) and $e^+e^-$ (solid)
colliders.  Updated by C.~Quigg from Ref.~\cite{Quigg:fy}.} \label{Top2002swa}
\end{figure}

Should we believe that the Higgs boson is light, as indicated by precision
electroweak data?  In defense of this, I show in Fig.~\ref{Top2002swa} a plot
of direct and indirect measurements of the top-quark mass {\it versus} time.
Precision electroweak measurements anticipated $m_t<200$ GeV, which suggests
that we should trust the prediction $m_h < 200$ GeV.

\newpage

\subsection{CKM}

The CKM elements involving the top quark have never been measured directly;
they are inferred from the unitarity of the CKM matrix.  Their values are
\cite{Hagiwara:pw}
\begin{eqnarray*}
|V_{td}|=0.004-0.014\\
|V_{ts}|=0.037-0.044\\
|V_{tb}|=0.9990-0.9993\;.
\end{eqnarray*}
Thus $|V_{td}|$, $|V_{ts}|$, and $|V_{tb}|$ are known with a precision of
$50\%$, $10\%$, and $0.02\%$, respectively.  How can we measure these CKM
elements?

\begin{figure}[b]
\begin{center}
\includegraphics[width=.45\textwidth]{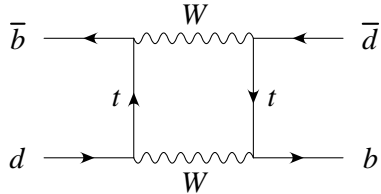}
\end{center}
\caption{$B_d^0-\bar B_d^0$ mixing proceeds via a box diagram.}
\label{bbbarmixing}
\end{figure}

\subsubsection{$|V_{td}|$} This may be determined indirectly from $B_d^0-\bar
B_d^0$ mixing, shown in Fig.~\ref{bbbarmixing}.  The frequency of oscillation,
$\Delta m_d$, is proportional to $|V_{tb}^*V_{td}|^2$.  Measurements give
\cite{Hagiwara:pw}
\begin{equation}
|V_{tb}^*V_{td}|=0.0079\pm 0.0015\;,
\end{equation}
where the uncertainty ($20\%$) is almost entirely from the theoretical
uncertainty in the hadronic matrix element.  Assuming three generations
($|V_{tb}|\approx~1)$, this is a more accurate measurement of $|V_{td}|$ than
can be inferred from unitarity ($50\%$).

\subsubsection{$|V_{ts}|$} This may be determined indirectly from $B_s^0-\bar
B_s^0$ mixing, which is the same as Fig.~\ref{bbbarmixing} but with the $d$
quark replaced by an $s$ quark. The frequency of oscillation, $\Delta m_s$, is
proportional to $|V_{tb}^*V_{ts}|^2$.  Thus far there is only a lower limit on
the oscillation frequency,
\begin{equation}
\Delta m_s > 14.4\, {\rm ps}^{-1}\;.
\end{equation}
The anticipated value from the range of $|V_{ts}|$ listed above is $\Delta
m_s\approx 18\, {\rm ps}^{-1}$, just above the current lower bound.  This
should be observable in Run II at the Tevatron.  However, the theoretical
uncertainty is very similar to that of $\Delta m_d$, which means that
$|V_{ts}|$ can only be extracted with an uncertainty of $20\%$, which is
greater than the uncertainty in the value inferred from unitarity ($10\%$).

\subsubsection{$|V_{ts}|/|V_{td}|$} We can use the similarity in the hadronic
matrix elements involved in $\Delta m_s$ and $\Delta m_d$ to our advantage by
taking the ratio:
\begin{equation}
\frac{\Delta m_s}{\Delta
m_d}=\frac{M_{B_s}}{M_{B_d}}\xi^2\left|\frac{V_{ts}}{V_{td}}\right|^2\;.
\end{equation}
The theoretical uncertainty in the ratio of the hadronic matrix elements,
$\xi^2$, is much less than the uncertainty in the hadronic matrix elements
themselves.  Using the value of $|V_{ts}|$ from unitarity yields an
uncertainty in $|V_{td}|$ that is less than the uncertainty obtained from
$\Delta m_d$ alone.

\begin{figure}[t]
\begin{center}
\includegraphics[width=.6\textwidth]{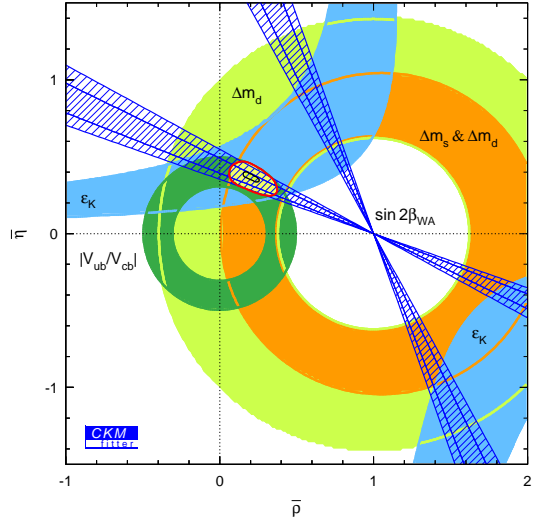}
\end{center}
\caption{The $\bar\rho$-$\bar\eta$ plane, showing constraints from various
measurements, as well as the best fit.  From http://ckmfitter.in2p3.fr
\cite{Hocker:2001xe}.} \label{rhoeta}
\end{figure}

I show in Fig.~\ref{rhoeta} the $\bar\rho$-$\bar\eta$ plane.  The radius of the
large circles centered at $(1,0)$ is proportional to $|V_{td}|$.  The large
annulus is from the measurement of $\Delta m_d$.  The small annulus that lies
inside it is from the ratio $\Delta m_s/\Delta m_d$ and $\Delta m_d$ combined,
using the current lower bound on $\Delta m_s$.  The measurement of $\Delta
m_s$ at the Tevatron will reduce the width of this annulus by about a half,
making it one of the most precise measurements in the $\bar\rho$-$\bar\eta$
plane.

\subsubsection{$|V_{tb}|$} Despite the fact that it has never been measured
directly, $|V_{tb}|$ is the best known CKM element ($0.02\%$), assuming three
generations.  It is only interesting to measure it if we relax the assumption
of three generations, in which case $|V_{tb}|$ is almost completely
unconstrained \cite{Hagiwara:pw},
\begin{equation}
|V_{tb}|=0.08-0.9993\;.
\end{equation}
How can we directly measure $|V_{tb}|$ in this scenario?

\begin{figure}[t]
\begin{center}
\includegraphics[width=.28\textwidth]{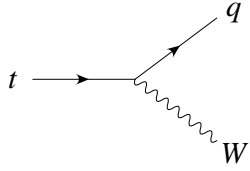}
\end{center}
\caption{Top-quark decay to a $W$ boson and a light quark ($q=d,s,b$).}
\label{topdecay}
\end{figure}

Let's consider top-quark decay, shown in Fig.~\ref{topdecay}.  CDF has measured
the fraction of top decays that yield a $b$ quark \cite{Affolder:2000xb},
\begin{eqnarray}
\frac{BR(t\to Wb)}{BR(t\to Wq)} &&= 0.94^{+0.31}_{-0.24}\nonumber\\
&&=\frac{|V_{tb}|^2}{|V_{td}|^2+|V_{ts}|^2+|V_{tb}|^2}\;,
\end{eqnarray}
where $q$ denotes any light quark ($d,s,b$).  The second line is the
interpretation of this measurement in terms of CKM elements.  If we were to
assume three generations, the denominator of this expression would be unity,
but we are not making that assumption.  The fact that this fraction is close to
unity only tells us that $|V_{tb}|\gg |V_{ts}|,|V_{td}|$; it does not tell us
its absolute magnitude.

\begin{figure}[b]
\begin{center}
\includegraphics[width=.8\textwidth]{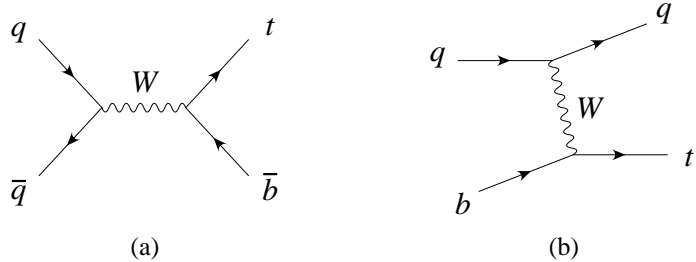}
\end{center}
\caption{Single-top-quark production via the weak interaction: (a) $s$-channel
process, (b) $t$-channel process.} \label{singletop}
\end{figure}

The way to measure $|V_{tb}|$ directly, with no assumptions about the number
of generations, is to measure top-quark production via the weak interaction
\cite{Stelzer:1998ni}. There are two relevant subprocesses at the Tevatron,
shown in Fig.~\ref{singletop}.  The $s$-channel process proceeds via
quark-antiquark annihilation and produces a $t\bar b$ final state.  Turning
this diagram on its side yields the $t$-channel process, in which the virtual
$W$ boson strikes a $b$ quark in the proton sea and promotes it to a top
quark.  Both processes produce a single top quark in the final state, rather
than a $t\bar t$ pair.  The cross sections for these processes are
proportional to $|V_{tb}|^2$, and thus provide a direct measurement of this
CKM element. These processes should be observed in Run II, and yield a
measurement of $|V_{tb}|$ with an uncertainty of about $10\%$.

\subsection{Top quark}

\begin{figure}[t]
\begin{center}
\includegraphics[width=.3\textwidth]{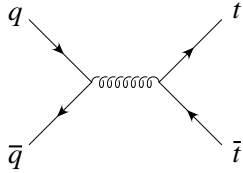}
\end{center}
\caption{Top-quark pair production at the Tevatron proceeds principally via
quark-antiquark annihilation, with a smaller contribution from $gg\to t\bar
t$.} \label{topproduction}
\end{figure}

The strong and weak interactions of the top quark are not nearly as well
studied as those of the other quarks and leptons.  The strong interaction is
most directly measured in top-quark pair production, shown in
Fig.~\ref{topproduction}. The weak interaction is measured in top-quark decay,
Fig.~\ref{topdecay}, and single-top-quark production, Fig.~\ref{singletop},
discussed in the previous section.

The standard model predicts that the $W$ boson in top-quark decay will be
dominantly longitudinally polarized,
\begin{equation}
BR(t\to W_0b) = \frac{m_t^2}{m_t^2+2M_W^2} \approx 0.70\;.
\end{equation}
CDF has made the crude measurement $BR(t\to W_0b)=0.91\pm 0.37 \pm 0.13$,
consistent with the standard-model expectation \cite{Affolder:1999mp}. This
measurement should improve significantly in Run II.

\begin{figure}[t]
\begin{center}
\includegraphics[width=.77\textwidth]{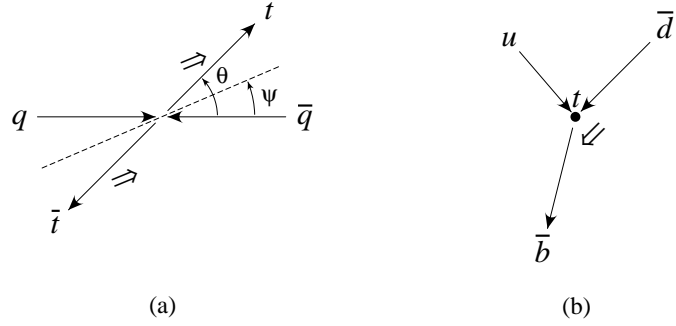}
\end{center}
\caption{(a) In $t\bar t$ production, the spins of the top quark and antiquark
are $100\%$ correlated when measured along an axis that makes an angle $\psi$
with respect to the beam axis, where $\tan
\psi=\beta^2\sin\theta\cos\theta/(1-\beta^2\sin^2\theta)$; (b) In single-top
production, the top quark is $100\%$ polarized along the direction of motion of
the $d$ quark, in the top-quark rest frame.} \label{topspin}
\end{figure}

One of the unique features of the top quark is that it decays before there is
time for its spin to be depolarized by the strong interaction.  Thus the
top-quark spin is directly observable via the angular distribution of its
decay products.  This means that we should be able to measure observables that
depend on the top-quark spin.

In the production of $t\bar t$ pairs via the strong interaction,
Fig.~\ref{topproduction}, the spins of the top quark and antiquark are $100\%$
correlated, as shown in Fig.~\ref{topspin}(a) \cite{Mahlon:1997uc}.  In
single-top production, the spin of the top quark is $100\%$ polarized along
the direction of motion of the $d$ quark, in the top-quark rest frame, as
shown in Fig.~\ref{topspin}(b) \cite{Mahlon:1996pn}.  These spin effects should
be observable in Run II.

\newpage

\subsection{Higgs boson}

\begin{figure}[b]
\begin{center}
\includegraphics[width=.8\textwidth]{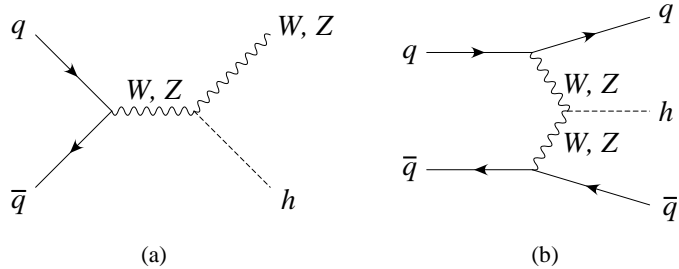}
\end{center}
\caption{Higgs-boson production (a) in association with a weak vector boson,
(b) via weak-vector-boson fusion.} \label{higgsproduction}
\end{figure}

As mentioned in the introduction to this section, the LHC was designed to
discover the Higgs boson.  However, there is a chance that the Tevatron could
make this discovery first.  The most promising channel is associated
production of the Higgs boson with a $W$ or $Z$ boson, as shown in
Fig.~\ref{higgsproduction}(a), followed by $h\to b\bar b$ \cite{Stange:ya}.
This channel is not considered viable at the LHC, so it is of particular
interest to try to observe it at the Tevatron. Other discovery channels are
associated production of the Higgs boson and a $t\bar t$ pair
\cite{Goldstein:2000bp} and, for higher Higgs-boson masses, Higgs-boson
production via gluon fusion, followed by $h\to WW^{(*)}$ \cite{Han:1998ma}.

\begin{figure}[t]
\begin{center}
\includegraphics[width=.77\textwidth]{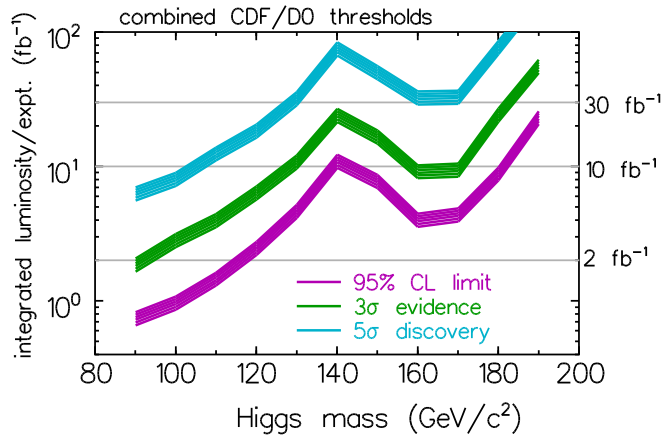}
\end{center}
\caption{Integrated luminosity required to discover ($5\sigma$), find evidence
for ($3\sigma$), or rule out ($95\%$ CL) the Higgs boson at the Tevatron {\it
vs.}~the Higgs mass.} \label{higgstevatron}
\end{figure}

Figure \ref{higgstevatron} is the well-known plot of the integrated luminosity
required to discover the Higgs boson ($5\sigma$), find evidence for it
($3\sigma$), or rule it out ($95\%$ CL) \cite{Carena:2000yx}.  If we take
seriously the indication from precision electroweak physics that the Higgs
boson is lighter than 200 GeV, then we will be unable to rule it out.  To
discover it, we will really need $5\sigma$ --- this is the Higgs boson we are
talking about! Given the lower bound $m_h>114$ GeV from LEP, this means we
need 15 fb$^{-1}$ of integrated luminosity to get into the game.

\begin{figure}[t]
\begin{center}
\includegraphics[width=.6\textwidth]{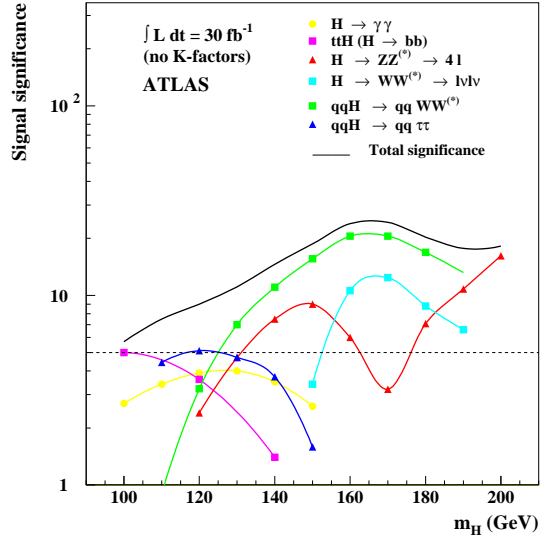}
\end{center}
\caption{$S/\sqrt B$ for a variety of Higgs-boson production and decay
channels at the LHC {\it vs.}~the Higgs mass.} \label{higgs_vbf_30}
\end{figure}

The LHC can't miss the Higgs boson.  In addition to the production processes
available at the Tevatron, there is also weak-vector-boson fusion, shown in
Fig.~\ref{higgsproduction}(b), which was originally proposed for the discovery
of a heavy Higgs boson. The potential of this process for the discovery of an
intermediate-mass Higgs boson has recently been appreciated
\cite{Rainwater:1998kj,Rainwater:1999sd}. Figure~\ref{higgs_vbf_30} shows the
signal significance for the discovery of an intermediate-mass Higgs boson in a
variety of channels.  The importance of the weak-vector-boson-fusion channels
is evident.

Once we discover the Higgs boson, we want to measure its couplings to other
particles.  The ratio of Higgs couplings can be extracted by measuring a
variety of production and decay modes \cite{Zeppenfeld:2000td}.  Thus it is
important to be able to see the Higgs boson in as many channels as possible.
This again emphasizes the importance of the weak-vector-boson-fusion channels.

\subsection{QCD}

Perhaps it is surprising to see QCD in this list of ways in which the Tevatron
can confront the standard model.  After all, we know beyond the shadow of a
doubt that QCD is the correct theory of the strong interaction.  However, we
don't always know how to use it correctly.  Here I discuss three aspects of the
confrontation between theory and experiment in QCD.

\subsubsection{$J/\psi$ production}

\begin{figure}[b]
\begin{center}
\includegraphics[width=.8\textwidth]{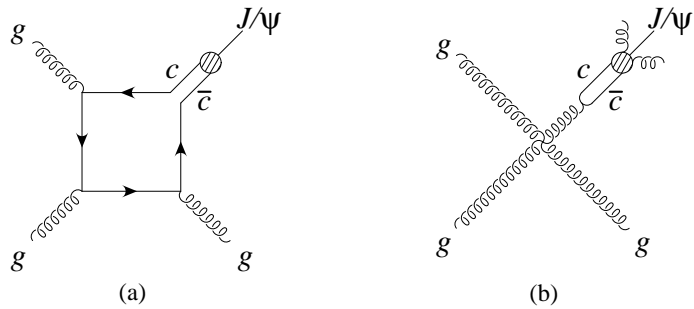}
\end{center}
\caption{$J/\psi$ production at high transverse momentum: (a) color-singlet
mechanism, (b) color-octet mechanism.} \label{jpsiproduction}
\end{figure}

\begin{figure}[t]
\begin{center}
\includegraphics[width=.7\textwidth]{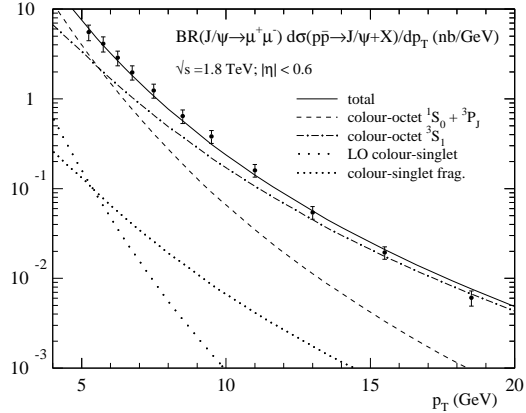}
\end{center}
\caption{$J/\psi$ transverse momentum distribution at the Tevatron. The LO
color-singlet contribution corresponds to Fig.~\ref{jpsiproduction}(a); the
color-octet $^3S_1$ contribution corresponds to Fig.~\ref{jpsiproduction}(b).
From Ref.~\cite{Kramer:2001hh}.} \label{psipt}
\end{figure}

\begin{figure}[b]
\begin{center}
\includegraphics[width=.6\textwidth]{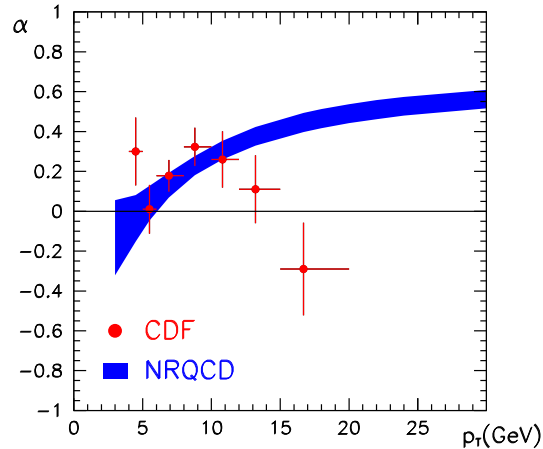}
\end{center}
\caption{Polarization of $J/\psi$ as a function of its transverse momentum. The
data are compared with the prediction from the process in
Fig.~\ref{jpsiproduction}(b) and related processes.} \label{psipol}
\end{figure}

It was long thought that $J/\psi$ production at high transverse momentum
proceeds via the process shown in Fig.~\ref{jpsiproduction}(a), in which a
color-singlet $c\bar c$ pair is produced.  However, this process yields a
cross section that is more than an order of magnitude too small, and has the
wrong transverse-momentum dependence, as shown by the curve labeled ``LO
color-singlet'' in Fig.~\ref{psipt}. We now believe that the dominant
production mechanism at high transverse momentum involves a gluon that produces
a color-octet $c\bar c$ pair, which then fragments into a $J/\psi$ by emitting
two or more soft gluons, as shown in Fig.~\ref{jpsiproduction}(b)
\cite{Braaten:1994vv}. For a suitable choice of the hadronic matrix element
that parameterizes the fragmentation function, this gives a good description
of the data, as shown in Fig.~\ref{psipt}. This mechanism also makes an
unambiguous prediction: the $J/\psi$ should be transversely polarized at high
transverse momentum, since it is emanating from a gluon, which has only
transverse polarization states. The predicted polarization is shown in
Fig.~\ref{psipol}, along with data from Run I; the agreement is hardly
encouraging \cite{Affolder:2000nn}. The data from Run II will put this to a
decisive test, and tell us if we need to go back to the drawing board in order
to explain $J/\psi$ production at high transverse momentum.

\subsubsection{$b$ production}

\begin{figure}[t]
\begin{center}
\includegraphics[width=.6\textwidth]{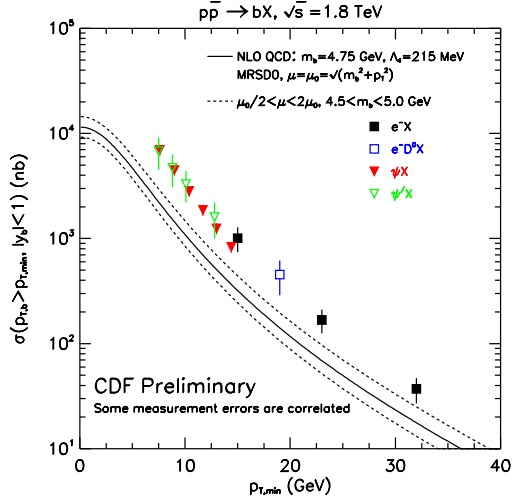}
\end{center}
\caption{The $b$-quark cross section {\it vs.} the minimum transverse momentum
of the $b$ quark.  From
http://www-cdf.fnal.gov/physics/new/bottom/bottom.html.} \label{bcrosssection}
\end{figure}

Another area in which the agreement between theory and experiment is less than
impressive is $b$-quark production.  The data lie significantly above the
prediction of next-to-leading-order QCD \cite{Nason:1989zy,Beenakker:1988bq},
as shown in Fig.~\ref{bcrosssection}. A similar excess has been observed at
HERA ($\gamma^*g\to b\bar b$) and at LEP ($\gamma^*\gamma^* \to b\bar b$).
This has stimulated a great deal of work on QCD, but there is not yet a
generally-accepted resolution of this conundrum. The slightly-higher energy of
Run II provides another venue in which to test our theoretical ideas against
experiment.

\subsubsection{Multijet production}

Multijet production is interesting in its own right, and also as a background
to new physics.  In order to recognize physics beyond the standard model, we
must first be able to calculate standard-model processes with reasonable
precision.  Multijet production (by itself or together with other particles)
is difficult to calculate even at tree level, and challenges our ability to
calculate efficiently.

There are two new general-purpose tools available for such calculations:
ALPGEN \cite{Mangano:2002ea} and MadEvent \cite{Maltoni:2002qb}.  They are
both leading-order event generators, with color information that allows them
to be merged with shower Monte Carlo codes.  Each has its own strengths, so I
invite you to take them out for a test drive.\footnote{ALPGEN:
http://mlm.home.cern.ch/mlm/alpgen \\ MadEvent:
http://madgraph.physics.uiuc.edu} Currently neither code can produce generic
multijet events with more than six jets, but that should change in the near
future. One promising new development in the calculation of multiparton
amplitudes is the color-flow decomposition; the subprocess cross section for
$gg\to 10 g$ has recently been evaluated using this method
\cite{Maltoni:2002mq}.

\section{Beyond the Standard Model}

As discussed in Section~\ref{sec:sm}, the standard model is a simple, elegant,
and successful theory.  Why should we even contemplate physics beyond the
standard model?  It almost seems ungrateful.  The situation is in some ways
analogous to that of classical physics in the 1890's, which seemed to explain
almost all known phenomena.  However, that decade witnessed the discovery of
X-rays by R\"ontgen, radioactivity by Becquerel, and the electron by Thomson,
and it soon became clear that these could not be understood in terms of
classical physics.  These discoveries and others eventually led to the
theories of quantum mechanics and relativity, upon which the standard model is
based.  The question I would like to address is: What are the analogous
anomalies today?  Below I list both direct and indirect evidence for physics
beyond the standard model.

\subsection{Direct evidence}

\subsubsection{Neutrinos}
\label{sec:numasses}

Recall that in our review of the standard model we restricted ourselves to the
simplest possible interactions, but we did not explain why we imposed this
restriction.  The most conservative (and most plausible) explanation is that
additional interactions are present, but they are suppressed.  If we were to
add these additional interactions, we would find (via dimensional analysis)
that they have coefficients with dimensions of an inverse power of mass
\cite{Willenbrock:2002ta},
\begin{equation}
{\cal L}={\cal L}_{SM}+\frac{1}{M}{\cal L}_5+\frac{1}{M^2}{\cal L}_6+\cdots
\end{equation}
where $M$ is a mass scale greater than the Higgs-field vacuum-expectation
value, $v$.  At energies much less than $M$, the least-suppressed interactions
come from the Lagrangian labeled ${\cal L}_5$.  Remarkably, there is one and
only one possible term in this Lagrangian (assuming the standard-model
particle content) \cite{Willenbrock:2002ta,Weinberg:sa},
\begin{equation}
{\cal L}_5 = c^{ij}(L_L^{iT}\epsilon\phi)C(\phi^T\epsilon L_L^j) +
h.c.\;.\label{dim5}
\end{equation}
where $L_L$ and $\phi$ are the lepton and Higgs-doublet fields (see
Table~\ref{tab:sm}) [$C$ is the charge-conjugation matrix].  When the Higgs
field acquires a vacuum-expectation value, this term gives rise to a Majorana
mass matrix for the neutrinos,
\begin{equation}
M_\nu^{ij}=c^{ij}\frac{v^2}{M}.
\end{equation}
Thus we expect neutrino masses and mixing, with masses much less than $v$ (for
$M\gg v$).  The observation of neutrino oscillations is thus unambiguous
evidence of physics beyond the standard model.

\subsubsection{Gravity}

Although we don't always think of it this way, gravity is definitely beyond
the standard model.  If we add a graviton field, $g_{\mu\nu}$, to the theory,
the least-suppressed additional interactions (using dimensional analysis) are
\begin{equation}
{\cal L}_{gravity} = \frac{M_P^2}{16\pi}\sqrt{-g}(-2\Lambda + R + \cdots)
\end{equation}
where $M_P$ is the Planck scale, $g\equiv {\rm det}\; g^{\mu\nu}$, $R$ is the
Ricci scalar, and $\Lambda$ is the cosmological constant.  The Ricci-scalar
term accounts for all of classical gravity.  The cosmological constant, long
thought to be exactly zero, is able to account for the mysterious ``dark
energy'' needed to accommodate cosmological observations.

\subsubsection{Astrophysics and Cosmology}

Along with the dark energy mentioned above, which is believed to account for
about $60\%$ of the mass-energy of the universe, there is also ``dark matter'',
whose nature is unknown, which accounts for about $35\%$ of the mass-energy.
Whatever this matter is, it is certainly beyond the standard model.  The
observed baryon asymmetry of the universe also cannot be explained by the
standard model, because it requires a source of CP violation beyond that
contained in the CKM matrix.  The inflationary model of the universe, so
successful in explaining many of the features of our universe, also requires
physics beyond the standard model.

\subsubsection{Precision electroweak}

As discussed in Section~\ref{sec:ew}, precision electroweak data may already
be indicating physics beyond the standard model.  Future measurements may
strengthen or weaken this evidence.

\subsection{Indirect evidence}

\subsubsection{Masses and mixing angles}

The standard model accommodates generic masses and mixing angles, but the
observed values are far from generic.  The natural scale of charged fermion
masses is of order $v$, but all charged fermions (except the top quark) are
much lighter than this, and display a hierarchical pattern.  The CKM mixing
angles are also not generic; they are small, and are also hierarchical.  These
facts suggest that there is physics beyond the standard model that explains the
pattern of charged fermion masses and mixing.  Unfortunately, the standard
model does not indicate at what energy scale this new physics resides
\cite{Maltoni:2001dc}.

\subsubsection{Grand Unification}

SU(5) grand unification is a lovely idea \cite{Georgi:sy}, but it is ruled
out; now that we know the gauge couplings with good accuracy, we find that
they do not unify at high energies. It is remarkable that by imposing
weak-scale supersymmetry on the theory, the relative evolution of the
couplings is nudged just enough to successfully unify the couplings at the
scale $M_{GUT}\approx 10^{16}$ GeV. This suggests that the supersymmetric
partners of the known particles await us as we probe the weak scale.

\subsubsection{Hierarchy problems}

Recall that the standard model has only one energy scale, the Higgs-field
vacuum-expectation value, $v$.  Why is $v\ll M_P, M_{GUT}$?  Why is $\Lambda\ll
M_P^2, v^2$?  In other words, why does it appear that physics beyond the
standard model is associated with scales wildly different from $v$? Perhaps
the explanation for this requires yet more physics beyond the standard model,
such as supersymmetry or large extra dimensions.

\subsection*{}

\indent\indent It is striking that almost all of these anomalies and hints of
physics beyond the standard model involve the Higgs field in one way or
another. Neutrino masses involve the Higgs field, via Eq.~(\ref{dim5}); the
vacuum-expectation value of the Higgs field contributes to the cosmological
constant; the axion (a type of Higgs field) is a dark-matter candidate; there
could be additional CP violation in the Higgs sector that generates the baryon
asymmetry; the inflaton (a scalar field) could drive inflation; precision
electroweak data constrain the Higgs sector; fermion masses and mixing angles
result from the coupling of the Higgs field to fermions, Eq.~(\ref{LYukawa});
SUSY SU(5) grand unification requires two Higgs doublets; and the hierarchy
problems involve the Higgs-field vacuum-expectation value.

The conclusion I would like to draw from these observations is that {\em
discovering and studying the Higgs boson (or bosons) is central to
understanding physics beyond the standard model}.  We should be on the lookout
for two (or more) Higgs doublets; Higgs singlets or triplets; Higgs-sector CP
violation; alternative models of electroweak symmetry breaking; composite
Higgs bosons, {\it etc}.  If we are lucky, we will begin to probe this physics
at the Tevatron; we are guaranteed to learn something about it at the LHC.  We
look forward to an exciting era of physics associated with electroweak
symmetry breaking and physics beyond the standard model.

\section*{Acknowledgements}

I would like to thank Thomas M\"uller for hosting a stimulating conference.  I
am grateful for assistance from U.~Baur, E.~Braaten, R.~K.~Ellis, W.~Giele,
A.~de Gouvea, A.~El-Khadra, T.~Liss, F.~Maltoni, W.~Marciano, U.~Nierste,
K.~Paul, K.~Pitts, C.~Quigg, D.~Rainwater, and T.~Stelzer.  This work was
supported in part by the U.~S.~Department of Energy under contracts
Nos.~DE-FG02-91ER40677 and DE-AC02-76CH03000.


%

\end{document}